# Identifying Supportive Student Factors for Mindset Interventions: A Two-model Machine Learning Approach

Nigel Bosch

School of Information Sciences and Department of Educational Psychology
University of Illinois Urbana–Champaign
pnb@illinois.edu

## Abstract

Growth mindset interventions foster students' beliefs that their abilities can grow through effort and appropriate strategies. However, not every student benefits from such interventions – yet research identifying which student factors support growth mindset interventions is sparse. In this study, we utilized machine learning methods to predict growth mindset effectiveness in a nationwide experiment in the U.S. with over 10,000 students. These methods enable analysis of arbitrarily-complex interactions between combinations of student-level predictor variables and intervention outcome, defined as the improvement in grade point average (GPA) during the transition to high school. We utilized two separate machine learning models: one to control for complex relationships between 51 student-level predictors and GPA, and one to predict the change in GPA due to the intervention. We analyzed the trained models to discover which features influenced model predictions most, finding that prior academic achievement, blocked navigations (attempting to navigate through the intervention software too quickly), self-reported reasons for learning, and race/ethnicity were the most important predictors in the model for predicting intervention effectiveness. As in previous research, we found that the intervention was most effective for students with prior low academic achievement. Unique to this study, we found that blocked navigations predicted an intervention effect as low as 0.185 GPA points (on a 0–4 scale) less than the mean. This was a notable negative prediction given that the mean intervention effect in our sample was just 0.026 GPA points, though few students (4.4%) experienced a substantial number of blocked navigation events. We also found that some minoritized students were predicted to benefit less (or even not at all) from the intervention. Our findings have implications for the design of computer-administered growth mindset interventions, especially in relation to students who experience procedural difficulties completing the intervention.

*Keywords: Data science applications in education, Secondary education, 21st century abilities*

# 1 Introduction

Students approach learning with differing beliefs about their own abilities to learn and grow (Dweck, 2006), beliefs about specific topics (Chestnut et al., 2018; Leslie et al., 2015), and reasons for learning (Yeager et al., 2014). These beliefs about learning, or *learning mindsets*, are related to learning outcomes. For example, mindsets can be a powerful influence on motivation, which is especially relevant to self-regulated learning with computers (Benson Soong et al., 2001; Robertson, 2011). Given this connection, recent research has explored the possibility of fostering certain learning mindsets to improve learning outcomes – often via student-oriented interventions (Aronson et al., 2002; Burnette et al., 2018; O'Rourke et al., 2014; Schmidt et al., 2017; Yeager et al., 2016). Such interventions sometimes improve student outcomes (e.g., grades), but are not effective for all students (Sisk et al., 2018; Yeager et al., 2019).

There are many types of mindsets students (and teachers) may adopt – e.g., mindset about learning with technology (Liu, 2011; Selim, 2007) – though the scope of all mindsets is too large for any one study. We focus specifically on students' growth mindset, which is the belief that one's ability can grow with appropriate effort and strategies (Dweck, 2006). Conversely, a fixed mindset reflects the belief that abilities and intelligence are more innate, and thus any impasses encountered while learning difficult concepts may be indicators of one's own lack of innate ability. Studies on the relationships between mindset and academic outcomes have produced mixed outcomes (Chao et al., 2017; Kaijanaho & Tirronen, 2018; Sisk et al., 2018; Warren et al., 2019). Similarly, some interventions designed to promote growth mindset have succeeded while others have not, and, surprisingly, interventions that successfully produced an increase in growth mindset were not successful at improving learning outcomes (Sisk et al., 2018). Understanding why mindset interventions are sometimes more or less helpful for different students is key to ensuring equitable benefit for students – especially traditionally-underserved students – by highlighting opportunities for refining interventions to suit the needs of particular sub-populations, and to avoid potential negative outcomes (e.g., wasting student time that could be better spent on a different educational activity).

Previous research has uncovered several reasons why growth mindset interventions produce mixed (sometimes even negative) academic outcomes. Growth mindset is hypothesized to relate to multiple constructs including effort, challenge-seeking, goal orientation, and success (Dweck, 2006), though recent research calls these relationships into question (Bahník & Vranka, 2017; Burgoyne et al., 2020). Nevertheless, interventions to promote growth mindset may be equally complex, promoting more than only growth mindset (Li & Bates, 2019). Thus, it is not always clear that the growth mindset aspect of an intervention is responsible for the results, particularly if multiple constructs are addressed in the intervention condition that are not in the control condition. In fact, in meta-analytic work, Sisk et al. (2018) found that growth mindset interventions only had positive effects on learning in the unexpected cases: when the interventions did not increase students' perceptions of the malleability of intelligence or when they did not measure that change. Sisk et al. also noted, however, that when growth mindset interventions succeeded in improving educational outcomes *on average*, the effects were inconsistent across demographic groups; students of low socioeconomic status (SES) were the only students to benefit. In this paper we focus on this last issue; when a growth mindset intervention is effective, on average but not for everyone, which students are likely to benefit?

Additionally, we explore a variety of additional constructs (e.g., stress, trust, challenge-seeking) that may explain intervention effects in addition to – or instead of – changes in mindset.

We utilize machine learning as a data-driven discovery method to explore many (potentially intersectional) student-level variables that characterize cases in which a large-scale computer-based mindset intervention is more or less beneficial. We examine the case of a computer-administered mindset intervention, which, combined with administrative records, affords analysis of many student-level variables including those related to their human–computer interaction behaviors during the intervention.

## 1.1 Computer-based mindset interventions

Computers are an attractive means of administering interventions directly to students at a wide scale, given their ubiquitous nature and flexibility. For example, the data we analyze in this study come from the National Study of Learning Mindsets (NSLM)[1], a large-scale computer-administered intervention experiment with over 10,000 students (Yeager et al., 2019). The NSLM experiment was, in turn, informed by an earlier computer-administered mindset intervention study with over 3,500 students in 10 schools in the United States (Yeager et al., 2016). In that study, they found that a growth mindset intervention improved 9th-grade grade point average (GPA), particularly for lower-achieving students. In fact, computer-based growth mindset interventions have been tested in a variety of educational settings, both formal and informal (Burnette et al., 2018; Donohoe et al., 2012; O'Rourke et al., 2014; Schmidt et al., 2017; Sisk et al., 2018; Yeager et al., 2016, 2019).

In one experiment, researchers modified a mathematics education game called *Refraction*, adding messaging designed to foster a growth mindset for students in the experimental condition, and adding neutral messaging about the importance of math in the control condition (O'Rourke et al., 2014). They found that the growth mindset messaging improved students' persistence (time spent) in the game. Similarly, researchers studying the growth mindset attitudes of students on a mobile computing learning platform found that growth mindset predicted higher quiz scores as well as longer time spent answering quizzes (Kizilcec & Goldfarb, 2019), though not in an experimental intervention context. These findings are consistent with the theory that growth mindsets help students persist in the face of adversity (Dweck, 2006). Furthermore, previous work has shown that students who struggle more are those who are most likely to benefit from persistence and challenge-seeking behaviors engendered by a growth mindset. For example, in the *Refraction* experiment, students who struggled most in the first few math problems later benefitted the most from the mindset intervention (O'Rourke et al., 2014). These findings also illustrate the fact that mindset intervention can be more or less likely to be effective depending on student characteristics and other contextual factors.

## 1.2 Heterogeneous intervention effects

Mindset intervention effects differ based on many factors, including cognitive factors (e.g., early struggles with math), demographic factors, and others. For example, O'Rourke et al. performed a follow-up *Refraction* study in which they compared mindset intervention effectiveness for students eligible for free/reduced-price school lunch, finding that their incentive-based mindset intervention was less effective (in terms of promoting engagement) for these students relative to their peers (O'Rourke et al., 2015).

---

[1] NSLM data access can be requested from mindset@prc.utexas.edu

In another study comparing achievement across SES levels, researchers found that growth mindset was correlated with SES and SES was correlated with scholastic achievement (Claro et al., 2016). However, the SES–achievement gap was mitigated for students from low-SES backgrounds when they had a growth mindset (i.e., the intervention was more effective for students from low-SES families in this case).

A recent meta-analysis of growth mindset interventions in education found that overall effects on academic outcomes are weak (Sisk et al., 2018); the overall effect size for interventions was $d = 0.080$ ($p = .010$). However, they also explored possible moderators of intervention effect. Student-level moderators included developmental stage (age), previous failed classes, experience of situational challenges (e.g., stereotype threat, moving to a new school), and eligibility for free/reduced-price school lunch. Procedural moderators included type of control condition (e.g., active, do-nothing), type of intervention (e.g., in class, computer administered), and timing of outcome measure (time to measurement of academic progress). They found that students with previous failed classes and students experiencing economic disadvantage benefitted significantly more than their peers.

In the NSLM experiment (Yeager et al., 2019), researchers utilized a computer-administered intervention to encourage adoption of a growth mindset for students in their first year of high school, a time when students are typically transitioning to a new school and may be encountering new challenges (Sisk et al., 2018). They found that students in lower-performing schools benefitted more from the intervention (compared to an active control group), in line with most previous research. However, there are many possible moderators (and combinations of moderators) that could potentially explain heterogeneity in intervention effects. Hence, in the current research we focus on methods to discover what these moderators are.

### 1.3 Data-driven discovery of moderators

Data-driven discovery refers to hypothesis generation from data (Romero et al., 2008). Applications of data-driven discovery include tasks like uncovering the structure of curricular topics (Méndez et al., 2014) or enumerating and measuring the cognitive states of students in computerized learning environments (Koedinger et al., 2013). In contrast, theory-driven analyses focus on relationships that have been hypothesized based on previous research. For example, in the primary analysis of the NSLM data, researchers hypothesized that students from lower-performing schools would benefit more from the intervention, given related findings in previous large-scale observational research (Claro et al., 2016).

Theory-driven analyses, in conjunction with appropriate experimental design methods like randomized controlled trials, provide a high degree of evidence for their results because relationships between variables are specified without seeing the data – thus avoiding overfitting conclusions to data. Data-driven approaches, on the other hand, can be prone to false positive results if large numbers of predictor variables are considered, model complexity is not sufficiently controlled (Ng, 2004), and cross-validation is not utilized (Picard & Cook, 1984). However, data-driven approaches allow discovery and measurement of heterogeneity that researchers had not or even could not have hypothesized in advance of data collection.

Data-driven discovery does not require defining specific hypotheses; rather, the choice of method and data defines a hypothesis space, which the method searches. In this paper, we utilize a type of machine learning model that defines a hypothesis space including individual moderators and combinations of moderators (see methods below). However, it is worth noting that no type of model is better, in general, than any other (known as the "no free lunch" theorem; Schaffer, 1994; Wolpert & Macready, 1997). Recent empirical review has shown that in practice the implications of the no free lunch theorem are often observed (Christodoulou et al., 2019). Moreover, simply because the hypothesis space defined by a method includes non-linear effects and complex moderators does not ensure they will be discovered if noise (e.g., measurement error) obscures these patterns (Jacobucci & Grimm, 2020). Thus, when answering the research questions described below, we also compared linear regression as a simple, well-known point of comparison (Appendix B).

## 1.4   Research questions

We expand on previous research by exploring a large number of possible student-level predictors (51 in total) gathered during the NSLM experiment, utilizing machine learning methods to predict intervention effectiveness on a per-student basis. The machine learning methods we employ here have notable advantages over simpler statistical analyses, especially for large datasets like the NSLM dataset. Of particular importance, they have the ability to handle arbitrarily-complex interactions between predictors, adjustable regularization methods to help prevent over-fitting of models, and suitable cross-validation strategies to measure accuracy appropriately even when over-fitting may have occurred. Machine learning models have proven successful for education-related tasks such as automatic test construction (El-Alfy & Abdel-Aal, 2008), personalized learning (Hsu et al., 2013; Lin et al., 2013), predicting student outcomes (Gray & Perkins, 2019; Lykourentzou et al., 2009), and studying student behaviors (Hew et al., 2020; Ninaus et al., 2019; Rico-Juan et al., 2019). In this study, we employ a novel approach that uses machine learning to discover moderators with complex relationships to heterogeneous intervention effects.

We had two objectives in training machine learning models. First, we sought to enable future improvements to the intervention by predicting when it would be less effective for improving post-intervention GPA (GPA at the end of the first year of high school). Second, recent advances in machine learning interpretability methods allow inspection of complicated models, thereby revealing which variables were most predictive. We were thus able to explore detailed predictors of intervention effectiveness that have never before been studied in such a large-scale randomized controlled trial (over 10,000 students from across the United States) and allow arbitrarily complex interactions between predictors (via machine learning methods). This is the first study to include procedural predictors such as self-reported distraction during the intervention administration, blocked navigation attempts, and others, alongside psychological measures of individual student differences. Blocked navigation events were defined as attempting to proceed through the intervention without spending at least 5 seconds on each page where a long reading was involved, or 10 seconds on each page where an open-ended text response was required. No minimum time restrictions were placed on short questions, such as those with multiple-choice answers.

We utilized the NSLM dataset to explore two research questions, which confirmed previous findings in the dataset and uncovered new findings with relevance to practical application of growth mindset interventions. Our research questions were:

RQ1) *How do student-level variables predict students' future GPA in the control condition?* By answering this question, we can estimate future GPA in absence of a mindset intervention, and thus control for this change for students in the intervention condition.

RQ2) *How do student-level variables predict intervention effectiveness (change in GPA due to intervention)?* In this analysis we explore whether the intervention works equally well for all students, and, if not, which student characteristics predict how well the intervention will work.

## 2   Material and methods

We answer the proposed research questions via a secondary analysis of the dataset from the National Study of Learning Mindsets, which was a double-blind randomized controlled trial testing a computer-administered growth mindset intervention. We registered methods based on analysis of a 10% random sample of data[2] (i.e., post-registration of methods), and analyses of the full dataset followed registered methods with two minor exceptions: 1) we used a slightly different machine learning algorithm (based on accuracy in only the initial 10% sample) and 2) we did not attempt a second cross-validation scheme described in the registration. The initial 10% sample was only examined to determine what types of variables were in the dataset (especially what types of missing data might occur) and to select the machine learning algorithm based on prediction accuracy measured via Pearson's $r$. All other analyses were performed after registration. Extensive details about the dataset and how it was collected are available in a publication from the initial analysis of the dataset (Yeager et al., 2019); in this section we summarize only key details relevant to the current analyses.

### 2.1   Participants

Participants were 16,310 students in 9th grade (the first year of high school), at 76 schools across the United States. Schools were chosen so that demographics were nationally representative. Of the 76 schools, 11 did not report administrative data required for the current analysis (e.g., student demographics). In the 65 schools that did report administrative data, students were 43% White, 24% Hispanic or Latinx, 11% Black or African American, 4% Asian American, and 18% other racial or ethnic groups. A further 3 of the 65 schools did not provide both pre- and post-experiment student grades for any students. Grades were required to construct the outcome measure in this paper, and these schools were therefore removed. Thus, there were 62 schools in the final sample. Grades were available for most students in these schools, but not all. There were 10,870 students with complete grade data that we analyzed in this study, consisting of 5,450 in the control condition and 5,420 in the intervention condition. For students who participated in the first term of 9th grade, pre-intervention GPA was defined as the mean of core course grades including math, English, social studies, and science from 8th grade (the last year of middle school) and post-intervention GPA was defined as the mean grades from core courses in both terms of 9th grade. For students who participated in the second term of 9th grade, pre-

---

intervention GPA was calculated from core courses in the first term of 9[th] grade and post-intervention GPA was calculated from the second term of 9[th] grade.

## 2.2  Intervention task

Students participated in the experiment in two 25-minute sessions, separated in time by one to four weeks depending on which school they attended. Both sessions included materials designed to encourage a growth mindset for the experimental condition, though different individual difference measures were administered at each session. The intervention consisted of messages and hypothetical scenarios intended to encourage growth mindset. For example, a message delivered to students near the beginning of the intervention read:

> High school is a time when the brain can learn and grow more than almost any other time in life. The work you do in high school can actually make your brain stronger, and building a stronger brain in high school helps you in life no matter what you plan to do.

In the control condition, students completed a similar computer-administered task, though without mention of growth mindset. Instead, students learned about how the brain works. An excerpt of material delivered to students near the beginning of the control activity read:

> Many people think the brain is a mystery. They don't know much about what it's made of, how it works, or what its different parts do.

The intervention was developed iteratively in previous work via qualitative and quantitative user-centered design (Yeager et al., 2016), with multiple rounds of focus groups, refinement, and testing. Some of the key features of the intervention included elements designed to 1) relate the intervention content specifically to first-year high school students, 2) promote relevance to students from community-oriented families and demographic groups, and 3) reinforce students' internalization of growth mindset by having them communicate its importance to a hypothetical future student.

During the intervention (and control) sessions, students completed a variety of individual difference measures. These included questions about their expectations for success in math during high school (Hulleman & Harackiewicz, 2009), their reasons for learning (Stephens et al., 2012; Yeager et al., 2014), their level of growth versus fixed mindset (Dweck, 2006), and others. The intervention software asked students procedural questions as well, such as whether they were distracted during the intervention, if nearby students were working diligently, if there were any technical difficulties, and related questions. The software also recorded information about students' interaction with the software, including how long they spent on each activity and how many times students' attempts to navigate within the software were blocked because they attempted to proceed too quickly without spending time on the material.

## 2.3  Machine learning procedure

Students activities and responses in the intervention task served as input variables to two machine learning models, one for each research question (Figure 1). The primary focus of this study is on RQ2: prediction of the intervention's effect on GPA from student-level variables. However, simply predicting change in GPA (i.e., post-experiment GPA – pre-experiment GPA) is not a suitable measure of intervention effect, because student GPAs may change in predictable ways even without the intervention. For example, students from low SES families may have a more (or less) difficult time transitioning from middle school to high school than their peers; thus, if we are interested in the relationship between SES and intervention effectiveness, we should first subtract the relationship between SES and GPA that occurs with no intervention. Hence, we first trained a model with data from only the control condition (*model 1*), and inspected this model to answer RQ1. Then, we applied model 1 to the intervention condition data, predicting how much each student's GPA in the intervention condition would change if they had not been given the intervention. Finally, we subtracted the predicted GPA change from the actual GPA change for each student in the intervention condition, yielding the residual GPA change due to the intervention after controlling for every predictor. These residual GPA changes served as labels (outcomes) to predict in the model for RQ2 (*model 2*), which was constructed and trained with the same procedure as model 1, detailed below.

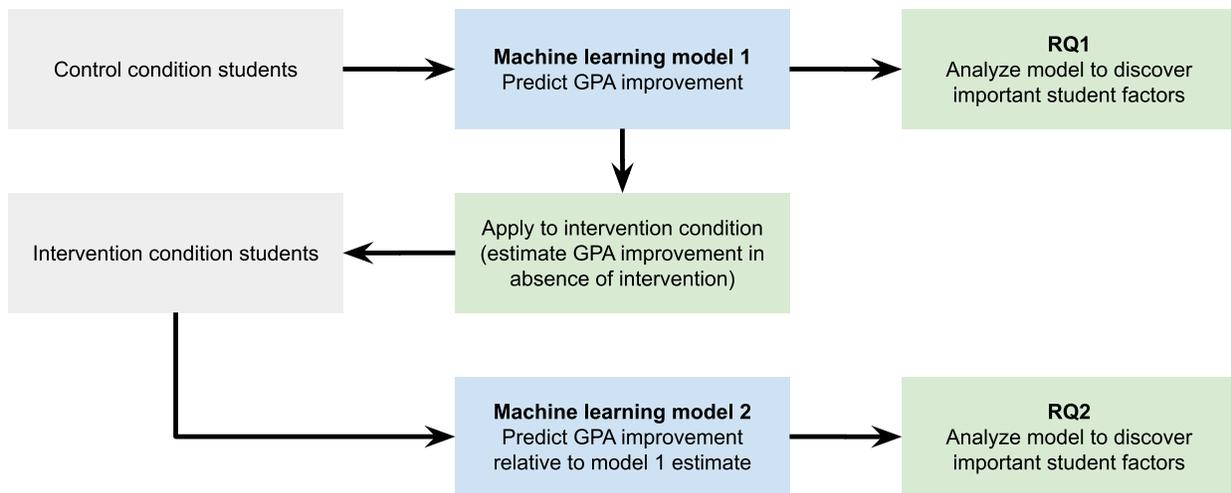

**Figure 1. Overview of the method in this study, illustrating how two machine learning models were applied to answer the two research questions in this study.**

### 2.3.1   Data preprocessing

The intervention software administered a battery of individual difference measures to students in both conditions. However, some measures were only provided to students in one condition. For example, we initially found the Field-specific Ability Beliefs (FAB) measure (Leslie et al., 2015) to be predictive of GPA change in the control condition; however, FAB was not administered in the intervention condition, so we removed it to maintain consistency across conditions. For the same reason, we removed measures of previous math experience, concerns about math student stereotypes, whether students were bored in math, perceptions of math teachers, and previous experience with growth mindset. A complete list of the final variables included (51 in total) is available in Appendix A. We trained a machine learning model to predict experimental condition from the final variables, which yielded non-significant predictive accuracy; this ensures that the final variable set did not reveal condition information that might unfairly influence results.

We also combined multiple variables into one for measures with multiple related questions. For example, we averaged two related questions about students' reasons for learning: whether they learn because they ultimately want to help others, and whether they learn to serve as a role model for others ($\alpha$ = .716). In a measure of challenge-seeking behavior, students were asked to create their own math worksheet from an array of easy, medium, and difficult exercises. We aggregated all of their responses into one variable consisting of the number of difficult minus the number of easy exercises they chose.

Values were missing in some cases, either because students did not respond to a question or because administrative data were unavailable. We handled these situations on a per-variable basis. For categorical variables, we added an additional category for missing values so that the models would be able to learn any important non-response patterns. For continuous and ordinal variables, we replaced missing values with the mean of that variable. In the case of ordinal variables, the means fell between ordinal ranks, thus effectively serving as a new rank and an indicator of missing values if missing-ness was predictive.

### 2.3.2   Model training

We trained gradient boosting tree models via the *XGBoost* package (T. Chen & Guestrin, 2016) in *R* (R Core Team, 2013); see Appendix B for a comparison to linear regression model accuracy. Gradient boosting is a powerful, state-of-the-art method for training models on high-level variables (structured data) such as the interpretable predictors in this study. Gradient boosting is a general framework that can be applied to learn different types of models by iteratively adding small sub-models to the overall model, where each new sub-model is specifically trained to address existing prediction errors in the previous sub-models. In this study, we chose to learn decision trees, which are well-suited to the current data because they allow for non-linear relationships, complex interactions between any number of predictors, and arbitrary distributions for each variable.

XGBoost has the capacity to fit decision boundaries as complicated as the training data itself, and thus can easily over-fit to training data. Therefore, we utilized leave-one-school-out cross-validation to evaluate prediction accuracy on held-out data (see Appendix B for a comparison to leave-several-schools-out), and tuned regularization hyperparameters to minimize over-fitting. Regularization methods constrain model complexity in various ways – for example, by enforcing an upper bound on the depth of trees in the model (i.e., degree of interaction) or enforcing a lower bound on the size of each leaf in each tree. We tuned hyperparameters including:

1) Maximum depth allowed for each tree in the model (1, 2, 3, …, 8)
2) Proportion of instances (rows) randomly sampled to train each tree (.5, .6, .7, …, 1)
3) Minimum number of instances required in each leaf of each tree (1, 2, 4, 8, …, 128)
4) Proportion of variables randomly sampled to train each tree (.5, .6, .7, …, 1)
5) Learning rate (.01, .02, .04, …, .32)
6) Minimum loss (error) reduction required to split a tree node (0, .01, .02, .04, …, 10.24)

We tuned hyperparameters with nested cross-validation in the training data only; i.e., we further split training data into train and validation subsets, trained models with various hyperparameter settings, and chose the best combination based on the validation subset. Given that there were 8 × 6 × 8 × 6 × 6 × 12 = 1,658,888 possible hyperparameter combinations to test for each of 62 cross-validation folds (one per school), we could not explore all hyperparameter combinations. Instead, we utilized coordinate descent (Wright, 2015). With this approach, we optimized each hyperparameter setting sequentially, in the order given above, iterating through the list of hyperparameters five times to allow for dependencies between hyperparameters to be partially resolved. However, even with the regularization included in this hyperparameter tuning method, findings may be particular to a specific dataset. Hence, we also repeated model 2 training with schools randomly divided into two subsamples, and found that results were similar across the two subsamples (see Appendix B for details).

We chose the best result based on lowest root mean squared error (RMSE) as measured via nested cross-validation within the training data[3]. There are many possible hyperparameter selection metrics, including combinations of multiple metrics (e.g., there are infinitely many variations of F-measure, which combines precision and recall metrics; Powers, 2011). Hyperparameter selection is an optimization problem, and thus there is no optimal choice of metric (Wolpert & Macready, 1997). However, RMSE is preferred in machine learning models of student outcomes (Pelánek, 2015; Sanyal et al., 2020), and since this is the same metric used by the model itself, we relied on RMSE. Finally, after selecting these hyperparameters we tuned the number of trees learned in the model, from 1 to 500 (based on training data only).

### 2.3.3  Model evaluation

Our research questions concern the variables that predict intervention effect, rather than estimation of the intervention effect itself (for which there are several recent machine learning methods; Athey & Wager, 2019; H. Chen et al., 2020; Microsoft Research, 2019). Thus, we inspected models to determine what they had learned about the relationships between potential predictors and GPA change. In particular, we calculated Shapley values for each feature (predictor) and instance (student). Shapley values measure the unique contribution of each feature toward the prediction of a particular instance (Lundberg & Lee, 2017), given all possible combinations of other features. In the regression models trained in this study, Shapley values can be easily interpreted as the difference in predicted outcome (i.e., difference in predicted GPA change) attributable to a particular feature. Shapley values are relative to the mean prediction, so mean Shapley values are always 0, thus limiting the feasibility of Shapley values for exploring overall directional effects; we instead examine mean absolute Shapley values as a measure of feature importance. Then, graphically examining Shapley values across all instances allows interpretation of non-linear effects after controlling for potential non-linear interactions with all other features. Shapley values can also be calculated for pairs of variables to assess specific interaction effects; however, given the number of variables in this analysis (51) and possible interactions of two variables ($51 \times 50 / 2 = 1,275$), we leave this for future work.

Note that Shapley values are in the same units as the outcome variable; thus, in this paper, Shapley values correspond to GPA points.

---

[3] The "best" model chosen according to lowest RMSE does not necessarily have statistically lower RMSE than all other models; rather, this is a common heuristic approach to choose a single model from among a set of candidates that may (or may not) be functionally equivalent.

# 3 Results

For each research question, we present model prediction accuracy to determine whether student-level variables significantly predict outcomes, then interpret models via Shapley values (in GPA units), which represent the key results.

## 3.1 RQ1: How do student-level variables predict students' future GPA in the control condition?

Mean GPA change in the control condition was -0.223 ($SD$ = 0.658, $n$ = 5,450), indicating that, on average, students in the control condition had lower post-experiment GPAs.

### 3.1.1 Model 1 prediction accuracy

We calculated $r$ for the students in each held-out school during cross-validation. Mean correlation between predicted and actual GPA change was $r$ = .327 ($p$ < .001), indicating that student-level variables significantly predicted changes in GPA in the control condition.

After measuring accuracy, we re-trained model 1 on all control condition data so that we could examine feature importance and apply the model to students in the experimental condition. We calculated aggregate feature importance as the mean of the absolute Shapley values (influence on prediction) across instances, given that we were interested in the size of influence rather than only influence in a specific direction. Given the large sample size, most variables had significantly above-zero influence on predictions; however, most were close to zero. We report the ten largest in Table 1; results for all variables are available in Appendix A.

### 3.1.2 Model 1 feature analysis

Results in Table 1 show, perhaps unsurprisingly, that past academic performance was the strongest indicator of future academic performance – both in terms of GPA as recorded in administrative records and in terms of student self-reports of their typical grades in core classes (English, math, and others). Similarly, expectations of future success in math classes indeed predicted future success. We graphically explored the relationships between the five most important variables and model 1 predictions below (Figure 2) to gather insight into possible directionality of effects. We focused only on the top five most important variables for brevity, given that the sixth-most important was notably less important (Table 1). In feature importance figures, variance in the $y$ direction for a particular value on the $x$ axis indicates an interaction with other variables; i.e., for instances with the same value in the $x$ variable the model made different predictions based on interactions with one or more additional features.

Figure 2 (upper left) shows that the relationship between pre-experiment GPA and predicted GPA was generally negative. Most notably, students with high pre-experiment GPAs (> 3; $n$ = 2,630) had little room for improvement, but could regress toward the mean; hence, predictions for these students ($M$ = -.181 relative to the mean) tended to be negative compared to their peers ($M$ = 0.166, $n$ = 2,820). Figure 2 (upper right) shows an opposite trend for students' self-reported typical grades. Predictions were higher (above zero GPA change) for students who reported getting "Mostly A's" ($M$ = 0.140 relative to the mean, $n$ = 2,112), while predicted GPA change was negative for students who reported low grades ($M$ = -0.089, $n$ = 3,340). Similarly, students' expectations for success in high school math were positively related to predicted GPA improvement (Figure 2 center left).

Analysis of predictions based on SES, as measured by eligibility for free or reduced-price lunch, showed higher predicted GPA improvement for students from high SES households (Figure 2 center right). Most notably, missing/not reported SES status was the most negative indicator of GPA change ($M = -0.104$ relative to the mean, $n = 2,010$; $M = 0.061$, $n = 3,440$ for their peers), though the influence on predictions was smaller for these less-important variables (as is apparent from the range of the $y$ axis). Gender had relatively little influence on the model (as is apparent from the range of the $y$ axis; Figure 2 bottom).

**Table 1. Feature importance (mean absolute Shapley values) for model 1. Means indicate the average predicted GPA change attributed to each feature; absolute values are reported because Shapley values refer to the difference from the mean prediction (hence mean Shapley value is always 0). Higher standard deviations indicate interactions with other variables.**

| Variable | Absolute effect on prediction in GPA points | |
|---|---|---|
| | *M* | *SD* |
| Pre-experiment GPA | 0.184 | 0.163 |
| Self-reported typical grades | 0.106 | 0.076 |
| Free or reduced-price lunch eligibility (SES) | 0.072 | 0.041 |
| Expectations of success in high school math (time 1) | 0.041 | 0.025 |
| Gender | 0.037 | 0.013 |
| Highest level of parental education | 0.024 | 0.018 |
| Expectations of success in high school math (time 2) | 0.024 | 0.014 |
| First year freshman | 0.022 | 0.020 |
| Race/ethnicity | 0.021 | 0.026 |
| Fixed mindset (time 2) | 0.021 | 0.014 |

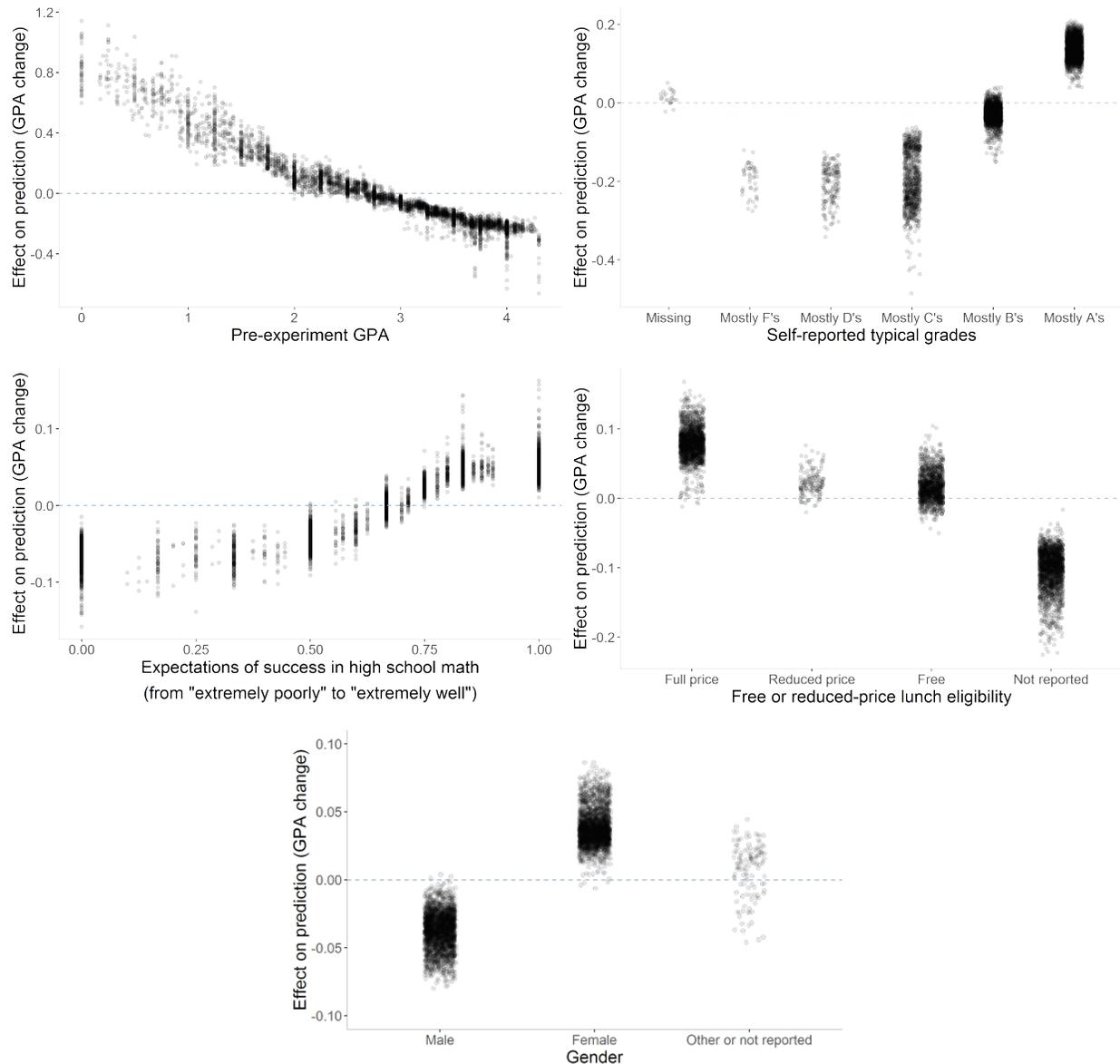

**Figure 2. Predicted GPA change attributed to the five most important variables in model 1, calculated via Shapley feature importance. A value of 0 (dashed line) indicates no difference relative to the mean prediction. The range of y-axis values varies considerably because some variables were much more important for prediction than others. Note that pre-experiment GPA (upper left) can be 0, e.g., for students with failing grades in schools with no-retention policies.**

### 3.2 RQ2: How do student-level variables predict intervention effectiveness (change in GPA due to intervention)?

Mean GPA change in the intervention condition was -0.197 ($SD$ = 0.674, $n$ = 5,420), indicating that students' GPAs decreased, on average, as was the case in the control condition ($M$ = -0.223, $SD$ = 0.658). However, the difference between GPA change in control and intervention conditions (0.026 GPA points; $d$ = 0.039) was significant ($p$ = .041), indicating that the intervention had a positive effect on GPA; for similar results focused on previously low-achieving students in the NSLM dataset, see Yeager et al., (2019). The small magnitude of the mean intervention effect on GPA is unsurprising given that the intervention was brief and was expected to benefit previously low-achieving students more than their peers. However, it serves to contextualize the findings here, since even small differences in intervention effect based on student-level variables may be large in terms of the overall intervention effectiveness.

We applied model 1 to the intervention condition to control for overall expected GPA change and GPA change related to individual differences. Mean predicted GPA in the intervention condition differed from actual GPA in the control condition by just 0.002, indicating that model 1 predicted similar GPAs for the intervention condition. The difference was not significant ($p$ = .873).

Residual difference between model 1's predicted GPA and actual GPA in the intervention condition was small ($M$ = 0.025, $SD$ = 0.674), as expected since the difference between conditions was small. However, even a small beneficial effect is valuable, given the brief nature of the intervention (50 minutes total). Moreover, variance suggests that there may be situations in which the intervention had a notably larger or smaller effect. Thus, we trained model 2 to predict these residuals attributable to the intervention.

### 3.2.1 Model 2 prediction accuracy
As in model 1 analysis, we calculated $r$ from held-out predictions during cross-validation for model 2. Mean $r$ was small (.094) but significantly above chance ($p$ < .001), indicating that the variables considered in this study predicted intervention efficacy even after controlling for their relationships to GPA in the control condition. Accuracy was low, which is expected given that the variables in this dataset explained only $R^2 = .327^2 = 10.7\%$ of variance in model 1, and model 2 is based on predictions from model 1.

### 3.2.2 Model 2 feature analysis
Shapley feature importance values in model 2 refer to the predicted change in GPA relative to expectations from the control condition (i.e., from model 1). Feature importance values for model 2 were smaller than those for model 1, since the mean of labels was closer to 0 and feature importance values are on the same scale as the labels. Feature importance results in Table 2 indicate that pre-experiment GPA was the most important predictor of the intervention effect ($M$ = 0.037, $SD$ = 0.028), despite controlling for overall relationships between past GPA and future GPA effects by applying model 1. This indicates that GPA was important for predicting the intervention effect itself. Also important was the count of blocked navigation events ($M$ = 0.019, $SD$ = 0.036), which are events where students attempted to proceed through the intervention software without answering required questions or without spending sufficient time reviewing intervention materials. The reason for learning variable ($M$ = 0.012, $SD$ = 0.007) refers to an individual difference measure of extrinsic purpose for learning (see Section 2.3.1). Student race/ethnicity ($M$ = 0.012, $SD$ = 0.005) was also among the most important variables. Previous research has also examined pre-intervention mindset (fixed vs. growth) as a moderator of intervention effect (Yeager et al., 2016); here, we found that it was not an important predictor ($M$ = 0.002, $SD$ = 0.002; see Table A1 in Appendix A), even though the intervention did produce a significant reduction in fixed mindset ($d$ = -0.223, $p$ < .001).

We explored non-linear feature importance effects, as for model 1, by comparing the effect each feature had on model 2 predictions across all students. For model 2, we examined only the top four most important features, given the notably smaller mean importance for the fifth-most important (Table 2). Figure 3 (upper left) shows the relationship between pre-experiment GPA and predicted intervention effect, highlighting that the predicted intervention effectiveness was higher for lower-GPA students, and was negative for higher-GPA students. Additionally, predictions varied more for lower-GPA students, indicating a greater degree of interaction with other variables.

The count of blocked navigation events shows a clear negative relationship with predicted intervention effect (Figure 3 upper right). There were few students (4.4% of intervention students, $n = 238$) who attempted a large number of blocked navigation events; however, for those who did, the predicted intervention effect was -0.185 GPA points relative to the mean. For their peers (95.6% of students, $n = 5,187$), predicted intervention effect was 0.008 GPA points above the mean.

Reason for learning had little effect on predictions. However, Figure 3 (lower left) shows that variance on the $y$ axis was larger for high values (i.e., when reason for learning was focused on being a role model and helping others) and low values than for intermediate values, which indicates interactions with other features for the high and low values of reason for learning. Race/ethnicity had similarly small effects, though what effect exists suggest that predictions were highest for White students ($M = 0.011$ relative to the mean, $n = 2390$), and lowest ($M = -0.022$, $n = 730$) for Black/African American students (Figure 3, lower right).

**Table 2. Mean absolute Shapley feature importance values for model 2 (predicting intervention effect). Time 1 and 2 refers to the first or second intervention session. Higher standard deviations indicate interactions with other variables.**

| Variable | Absolute effect on prediction | |
|---|---|---|
| | *M* | *SD* |
| Pre-experiment GPA | 0.037 | 0.028 |
| Blocked navigation count | 0.019 | 0.036 |
| Reason for learning | 0.012 | 0.007 |
| Race/ethnicity | 0.012 | 0.005 |
| Expectations of success in high school math (time 2) | 0.008 | 0.006 |
| Distraction during intervention (time 1) | 0.008 | 0.003 |
| Self-reported typical grades | 0.007 | 0.011 |
| School culture regarding effort in classes | 0.007 | 0.008 |
| Other students worked quietly on the intervention (time 1) | 0.007 | 0.011 |
| Math teacher is respectful | 0.007 | 0.009 |

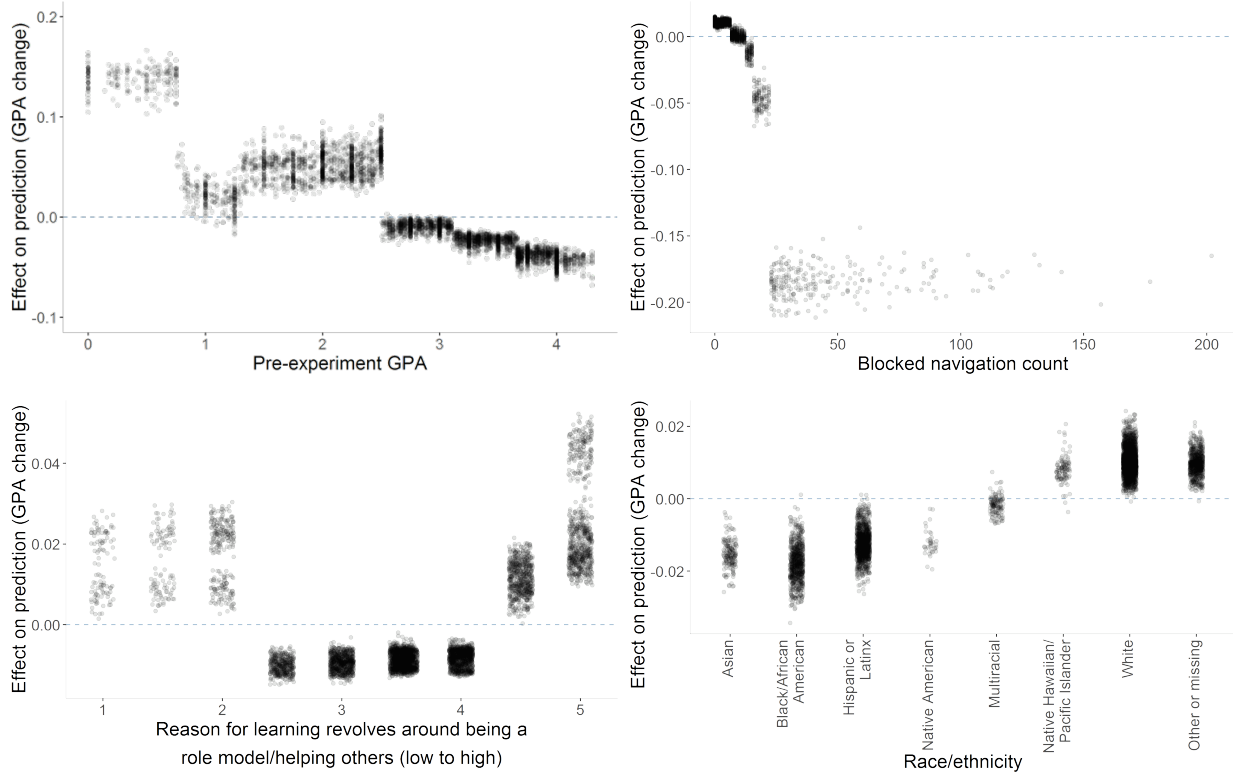

**Figure 3. Shapley feature importance values for the four most important variables in model 2. Most notably, the blocked navigation count (upper right) had a large effect on model predictions for students who experienced many such events.**

## 4   Discussion

In this study, we were interested in discovering possible indicators of when the NSLM intervention would work, including as many predictive variables as possible to identify indicators that have not been previously explored. We utilized machine learning methods to account for complex variable interactions and to perform automatic feature selection, and measure the importance of features to ascertain relationships between variables and predictions. In this section we discuss our main findings from these analyses, implications of these findings, and possibilities for future work.

### 4.1   Main findings

Our first research question focused on using feature importance analysis to discover what student-level characteristics predict grade point average (GPA) during the transition from middle school to high school. Given that students with high GPAs have comparatively little room for improvement, we expected that prior GPA would predict whether or not student GPA would improve in the control condition. This was indeed the case. In general, GPA decreased as students transitioned to high school, and students with higher GPAs (and little room to improve) were predicted to experience the largest GPA decreases. Related backfire effects were previously observed for students in high-achieving schools (Yeager et al., 2019). However, students' self-reported grades showed the opposite trend, and closely matched their expectations of success. This contrast between the effects of GPA and self-perceptions of grades suggests possible differences in students' metacognitive assessments of knowledge, where some students' self-perceptions of knowledge may be negatively related to their actual knowledge (Pennycook et al., 2017).

We also expected demographic differences in both predicted GPA change (research question 1/model 1) and predicted intervention effect (research question 2/model 2), given previous research on race, ethnicity, gender identity, SES, and other demographic factors (Akos & Galassi, 2004; Benner, 2011; Benner & Graham, 2009; Bian et al., 2018; Claro et al., 2016). In our results, SES and gender effects were notable for model 1 predictions, and race/ethnicity was for model 2. Directionality of these effects largely matches those in previous research, though the effect of SES in model 1 is notable. In particular, students for whom free/reduced-price lunch eligibility information was not available were predicted as having more negative GPA change than the other groups. This may be a function of systematic differences between schools, though it is unclear without additional data about those schools. Perhaps most importantly, demographic variables were not particularly strong predictors of intervention effectiveness according to model 2, compared to pre-experiment GPA and the blocked navigation count procedural measure.

Finally, feature importance analysis showed that fixed vs. growth mindset was one of the ten most important predictors in model 1 (Table 1), but not for model 2 (Table 2). In model 2, fixed mindset pre- and post-intervention were the 26th and 16th most important predictors (Table A1 in Appendix A), indicating that mindset and change in mindset (which a machine learning model could derive from the two timepoints) were not driving most of the change seen in students' GPA. This is despite the fact that the intervention reduced fixed mindset, as expected. Other factors, such as prior GPA and distractions encountered during the intervention were more crucial for predicting how much students' GPAs would change due to the intervention. This is in contrast to the simpler hypothesis that prior fixed mindset is a key moderator (Yeager et al., 2016), but in accord with work suggesting the effects are moderated by many factors (Chao et al., 2017) and that non-mindset constructs like effort encouragement may drive intervention efficacy (Li & Bates, 2019). In sum, these findings highlight the complexity of mindset interventions and the difficulty of attributing overall intervention results to specific constructs in the presence of many confounding variables.

### 4.2 Implications for computer-administrated learning mindset interventions

Our analyses focus primarily on the effects of input variables on model predictions, rather than model accuracy. If prediction accuracy for model 2 was nearly perfect, the model could, in theory, be applied at the individual student level to identify those for whom the intervention would provide no benefit, and thus avoid wasting student time and computing resources on the intervention. In practice, however, prediction accuracy was low ($r = .094$, $p < .001$), which is expected given the inherent difficulty of predicting a small intervention effect ($d = 0.039$, $p = .041$). Although small intervention effects are expected in large-scale educational research (Kraft, 2020), low prediction accuracy is still a large limitation since it indicates there is significant unexplained variance accounted for by unobserved variables (e.g., emotions, metacognition, life events).

However, there are implications that can be drawn from model 2 (though note that implications are limited to this specific intervention, which encourages multiple changes including adopting a growth mindset and appropriate strategies; see Introduction). First, the intervention appeared, in general, to be more effective for lower-achieving students, which aligns with previous work based on other methods and datasets (Sisk et al., 2018; Yeager et al., 2019). Thus, if a growth mindset intervention is to be targeted to a specific population, the best target is likely students with lower prior academic achievement. Second, demographic variables were not especially important for model 2 predictions. Notably, the strongest demographic predictor (race/ethnicity) influenced predictions less than 0.022 GPA points in either direction for each group. Thus, model 2 suggests that the intervention worked (if not necessarily equally well) across demographic groups. Third, the count of blocked navigation attempts was a predictor of ineffective interventions for some students. Few students experienced many blocked navigation attempts (4.4%), though predicted intervention effect was notably lower for those students (-0.185 GPA points relative to the mean; 0.008 for their peers). This result is especially crucial for administering computer-based mindset interventions, and merits future work with these students to discover their reasons for encountering blocked navigation attempts, and to develop methods that are more suited to their particular needs. For example, students may have encountered blocked navigation attempts because of instructions they found unclear, in which case instructions could be clarified or customized for those students. Alternatively, students may have attempted to rush through the intervention out of a dislike for extended computer-based interventions, in which case more research is needed to determine how to improve the intervention software (or avoid it altogether for some students). Interventions may also benefit from increased teacher involvement if administered in classroom environments, specifically for the purpose of encouraging students to remain engaged with the intervention task.

Our findings also highlight the importance of selecting an appropriate imputation method for missing data when analyzing the intervention results. In particular, adding a unique category for missing data was key to the predictive value of the SES proxy variable (eligibility for free/reduced-price lunch). Strategies such as replacement with the mode or averaging across multiple branches of decision trees would diminish this effect.

## 4.3   Limitations and future work

There are a few limitations of the analyses in this paper that should be addressed in future work. First, the intervention effect (outcome) we focused on was limited to change in GPA. However, there may be other effects. For example, adopting a growth mindset may encourage students to seek more challenging coursework in the future (Yeager et al., 2019), thereby promoting learning but not necessarily improving GPA. Our analyses could be repeated in future work with alternative outcomes, such as future course-taking behaviors.

Second, the analysis of features in both machine learning models is exploratory in nature. While analyses were registered, specific hypotheses could not be registered in advance, thus limiting the utility of registration for avoiding false positive findings. It is possible that there are multiple explanations for the predictive patterns found by the models (e.g., noise), since accuracy was far from perfect. However, given that we employed multiple forms of regularization tuned via nested cross-validation, trained on a nationally-representative dataset, and found effects in line with previous research on the NSLM dataset, it appears likely that patterns captured by the models are not spurious. Future confirmatory analysis (rather than data-driven discovery) is especially needed to systematically explore the connection between blocked navigation events and lower predicted intervention effectiveness, since this is a novel finding with implications for computer-administered mindset interventions.

Third, this study focused on individual student-level predictors of intervention effectiveness. The Shapley feature analysis method we applied accounted for interactions between multiple variables, but did not provide insight into what those interactions were. Interpretability of Shapley analyses is also limited to graphical interpretation, given that effects can be non-linear and lack clear directionality. Moreover, there may be school-level predictors (context variables) that contribute to explaining variance in intervention effectiveness. For example, student-to-teacher ratio may relate to how much support students receive when they most need it, and race/ethnicity could be more important in some schools than other (e.g., when a student is a member of a minority group consisting of 2% of the student body versus 25% of the student body). Future work may thus offer additional insights by focusing on school-level predictors and theoretically-motivated interactions between predictors. However, context variables introduce multiple levels of dependency (school, teacher, and student), which machine learning methods are typically not equipped to recognize. Approaches like boosting may be required to avoid overfitting (Freund & Schapire, 1997); for example, a model might be trained first on school-level context variables, since there are relatively few schools (compared to teachers or students), then a second model could be trained on teacher-level context variables, followed by a student-level model.

Finally, the NSLM study included many measures that may not be part of intervention implementations in practice, and, as noted in Section 2.3.1, not all measures were administered to all students in both conditions. Future work is thus needed to determine whether measures included in the intervention software for research purposes contribute to intervention effects, given that these surveys may be absent in future implementations of the software.

## 4.4   Concluding remarks

Computer-administered growth mindset interventions can be beneficial for learning, as results have shown. However, not every student benefits. It is important to understand which students and groups of students benefit, to avoid inequitable outcomes and enable more judicious use of resources. Thus, in this study we considered a large number of possible predictors of intervention effectiveness and found that having a need for intervention (i.e., lower GPA) is important, but also that the mindset intervention was – unsurprisingly – ineffective when students experienced frequent navigation issues while interacting with the intervention software. Our findings will influence future research on procedural features, like blocked navigation attempts, during computer-administered mindset interventions.

# Acknowledgements

Research reported in this manuscript was supported by the National Study of Learning Mindsets Early Career Fellowship with funding generously provided by the Bezos Family Foundation to Student Experience Research Network (formerly Mindset Scholars Network) and the University of Texas at Austin Population Research Center. The University of Texas at Austin receives core support from the National Institute of Child Health and Human Development under the award number 5R24 HD042849.

This manuscript uses data from the National Study of Learning Mindsets (doi:10.3886/ICPSR37353.v1) (PI: D. Yeager; Co-Is: R. Crosnoe, C. Dweck, C. Muller, B. Schneider, & G. Walton), which was made possible through methods and data systems created by the Project for Education Research That Scales (PERTS), data collection carried out by ICF

International, meetings hosted by Student Experience Research Network at the Center for Advanced Study in the Behavioral Sciences at Stanford University, assistance from C. Hulleman, R. Ferguson, M. Shankar, T. Brock, C. Romero, D. Paunesku, C. Macrander, T. Wilson, E. Konar, M. Weiss, E. Tipton, and A. Duckworth, and funding from the Raikes Foundation, the William T. Grant Foundation, the Spencer Foundation, the Bezos Family Foundation, the Character Lab, the Houston Endowment, the National Institutes of Health under award number R01HD084772-01, the National Science Foundation under grant number 1761179, Angela Duckworth (personal gift), and the President and Dean of Humanities and Social Sciences at Stanford University.

The content is solely the responsibility of the author(s) and does not necessarily represent the official views of the Bezos Family Foundation, Student Experience Research Network, the University of Texas at Austin Population Research Center, the National Institutes of Health, the National Science Foundation, or other funders.

# References

Akos, P., & Galassi, J. P. (2004). Gender and race as variables in psychosocial adjustment to middle and high school. *The Journal of Educational Research*, *98*(2), 102–108. https://doi.org/10.3200/JOER.98.2.102-108

Aronson, J., Fried, C. B., & Good, C. (2002). Reducing the effects of stereotype threat on african american college students by shaping theories of intelligence. *Journal of Experimental Social Psychology*, *38*(2), 113–125. https://doi.org/10.1006/jesp.2001.1491

Athey, S., & Wager, S. (2019). Estimating treatment effects with causal forests: An application. *ArXiv:1902.07409 [Stat]*. http://arxiv.org/abs/1902.07409

Bahník, Š., & Vranka, M. A. (2017). Growth mindset is not associated with scholastic aptitude in a large sample of university applicants. *Personality and Individual Differences*, *117*, 139–143. https://doi.org/10.1016/j.paid.2017.05.046

Benner, A. D. (2011). The transition to high school: Current knowledge, future directions. *Educational Psychology Review*, *23*(3), 299–328. https://doi.org/10.1007/s10648-011-9152-0

Benner, A. D., & Graham, S. (2009). The transition to high school as a developmental process among multiethnic urban youth. *Child Development*, *80*(2), 356–376. https://doi.org/10.1111/j.1467-8624.2009.01265.x

Benson Soong, M. H., Chuan Chan, H., Chai Chua, B., & Fong Loh, K. (2001). Critical success factors for on-line course resources. *Computers & Education*, *36*(2), 101–120. https://doi.org/10.1016/S0360-1315(00)00044-0

Bian, L., Leslie, S.-J., Murphy, M. C., & Cimpian, A. (2018). Messages about brilliance undermine women's interest in educational and professional opportunities. *Journal of Experimental Social Psychology*, *76*, 404–420. https://doi.org/10.1016/j.jesp.2017.11.006

Burgoyne, A. P., Hambrick, D. Z., & Macnamara, B. N. (2020). How firm are the foundations of mind-set theory? The claims appear stronger than the evidence. *Psychological Science*, *31*(3), 258–267. https://doi.org/10.1177/0956797619897588

Burnette, J. L., Russell, M. V., Hoyt, C. L., Orvidas, K., & Widman, L. (2018). An online growth mindset intervention in a sample of rural adolescent girls. *British Journal of Educational Psychology*, *88*(3), 428–445. https://doi.org/10.1111/bjep.12192

Chao, M. M., Visaria, S., Mukhopadhyay, A., & Dehejia, R. (2017). Do rewards reinforce the growth mindset?: Joint effects of the growth mindset and incentive schemes in a field intervention. *Journal of Experimental Psychology: General*, *146*(10), 1402–1419. https://doi.org/10.1037/xge0000355

Chen, H., Harinen, T., Lee, J.-Y., Yung, M., & Zhao, Z. (2020). CausalML: Python package for causal machine learning. *ArXiv:2002.11631 [Cs, Stat]*. http://arxiv.org/abs/2002.11631

Chen, T., & Guestrin, C. (2016). XGBoost: A scalable tree boosting system. *Proceedings of the 22nd ACM SIGKDD International Conference on Knowledge Discovery and Data Mining*, 785–794. https://doi.org/10.1145/2939672.2939785

Chestnut, E. K., Lei, R. F., Leslie, S.-J., & Cimpian, A. (2018). The myth that only brilliant people are good at math and its implications for diversity. *Education Sciences*, *8*(2), 65. https://doi.org/10.3390/educsci8020065

Christodoulou, E., Ma, J., Collins, G. S., Steyerberg, E. W., Verbakel, J. Y., & Van Calster, B. (2019). A systematic review shows no performance benefit of machine learning over logistic regression for clinical prediction models. *Journal of Clinical Epidemiology*, *110*, 12–22. https://doi.org/10.1016/j.jclinepi.2019.02.004

Claro, S., Paunesku, D., & Dweck, C. S. (2016). Growth mindset tempers the effects of poverty on academic achievement. *Proceedings of the National Academy of Sciences*, *113*(31), 8664–8668. https://doi.org/10.1073/pnas.1608207113

Donohoe, C., Topping, K., & Hannah, E. (2012). The impact of an online intervention (Brainology) on the mindset and resiliency of secondary school pupils: A preliminary mixed methods study. *Educational Psychology*, *32*(5), 641–655. https://doi.org/10.1080/01443410.2012.675646

Dweck, C. S. (2006). *Mindset: The New Psychology of Success*. Random House.

El-Alfy, E.-S. M., & Abdel-Aal, R. E. (2008). Construction and analysis of educational tests using abductive machine learning. *Computers & Education*, *51*(1), 1–16. https://doi.org/10.1016/j.compedu.2007.03.003

Freund, Y., & Schapire, R. E. (1997). A decision-theoretic generalization of on-line learning and an application to boosting. *Journal of Computer and System Sciences*, *55*(1), 119–139. https://doi.org/10.1006/jcss.1997.1504

Gray, C. C., & Perkins, D. (2019). Utilizing early engagement and machine learning to predict student outcomes. *Computers & Education*, *131*, 22–32. https://doi.org/10.1016/j.compedu.2018.12.006

Hew, K. F., Hu, X., Qiao, C., & Tang, Y. (2020). What predicts student satisfaction with MOOCs: A gradient boosting trees supervised machine learning and sentiment analysis approach. *Computers & Education*, *145*, 103724. https://doi.org/10.1016/j.compedu.2019.103724

Hsu, C.-K., Hwang, G.-J., & Chang, C.-K. (2013). A personalized recommendation-based mobile learning approach to improving the reading performance of EFL students. *Computers & Education*, *63*, 327–336. https://doi.org/10.1016/j.compedu.2012.12.004

Hulleman, C. S., & Harackiewicz, J. M. (2009). Promoting interest and performance in high school science classes. *Science*, *326*(5958), 1410–1412. https://doi.org/10.1126/science.1177067

Jacobucci, R., & Grimm, K. J. (2020). Machine learning and psychological research: The unexplored effect of measurement. *Perspectives on Psychological Science*, *15*(3), 809–816. https://doi.org/10.1177/1745691620902467

Kaijanaho, A.-J., & Tirronen, V. (2018). Fixed versus growth mindset does not seem to matter much: A prospective observational study in two late bachelor level computer science courses. *Proceedings of the 2018 ACM Conference on International Computing Education Research*, 11–20. https://doi.org/10.1145/3230977.3230982

Kizilcec, R. F., & Goldfarb, D. (2019). Growth mindset predicts student achievement and behavior in mobile learning. *Proceedings of the Sixth (2019) ACM Conference on Learning @ Scale*, 8:1–8:10. https://doi.org/10.1145/3330430.3333632

Koedinger, K. R., Stamper, J. C., McLaughlin, E. A., & Nixon, T. (2013). Using data-driven discovery of better student models to improve student learning. In H. C. Lane, K. Yacef, J. Mostow, & P. Pavlik (Eds.), *Artificial Intelligence in Education* (pp. 421–430). Springer. https://doi.org/10.1007/978-3-642-39112-5_43

Kraft, M. A. (2020). Interpreting effect sizes of education interventions. *Educational Researcher*, *49*(4), 241–253. https://doi.org/10.3102/0013189X20912798

Leslie, S.-J., Cimpian, A., Meyer, M., & Freeland, E. (2015). Expectations of brilliance underlie gender distributions across academic disciplines. *Science*, *347*(6219), 262–265. https://doi.org/10.1126/science.1261375

Li, Y., & Bates, T. C. (2019). You can't change your basic ability, but you work at things, and that's how we get hard things done: Testing the role of growth mindset on response to setbacks, educational attainment, and cognitive ability. *Journal of Experimental Psychology: General*, *148*(9), 1640. https://doi.org/10.1037/xge0000669

Lin, C. F., Yeh, Y., Hung, Y. H., & Chang, R. I. (2013). Data mining for providing a personalized learning path in creativity: An application of decision trees. *Computers & Education*, *68*, 199–210. https://doi.org/10.1016/j.compedu.2013.05.009

Liu, S.-H. (2011). Factors related to pedagogical beliefs of teachers and technology integration. *Computers & Education*, *56*(4), 1012–1022. https://doi.org/10.1016/j.compedu.2010.12.001

Lundberg, S. M., & Lee, S.-I. (2017). A unified approach to interpreting model predictions. In I. Guyon, U. V. Luxburg, S. Bengio, H. Wallach, R. Fergus, S. Vishwanathan, & R. Garnett (Eds.), *Advances in Neural Information Processing Systems 30* (pp. 4765–4774). Curran Associates, Inc.

Lykourentzou, I., Giannoukos, I., Nikolopoulos, V., Mpardis, G., & Loumos, V. (2009). Dropout prediction in e-learning courses through the combination of machine learning techniques. *Computers & Education*, *53*(3), 950–965. https://doi.org/10.1016/j.compedu.2009.05.010

Méndez, G., Ochoa, X., & Chiluiza, K. (2014). Techniques for data-driven curriculum analysis. *Proceedings of the Fourth International Conference on Learning Analytics And Knowledge*, 148–157. https://doi.org/10.1145/2567574.2567591

Microsoft Research. (2019). *EconML: A Python package for ML-based heterogeneous treatment effects estimation* (0.x) [Computer software]. https://github.com/microsoft/EconML

Ng, A. Y. (2004). Feature selection, L1 vs. L2 regularization, and rotational invariance. *Proceedings of the Twenty-First International Conference on Machine Learning (ICML 2004)*, 78–85. https://doi.org/10.1145/1015330.1015435

Ninaus, M., Greipl, S., Kiili, K., Lindstedt, A., Huber, S., Klein, E., Karnath, H.-O., & Moeller, K. (2019). Increased emotional engagement in game-based learning – A machine learning approach on facial emotion detection data. *Computers & Education*, *142*, 103641. https://doi.org/10.1016/j.compedu.2019.103641

O'Rourke, E., Chen, Y., Haimovitz, K., Dweck, C. S., & Popović, Z. (2015). Demographic differences in a growth mindset incentive structure for educational games. *Proceedings of the Second (2015) ACM Conference on Learning @ Scale*, 331–334. https://doi.org/10.1145/2724660.2728686

O'Rourke, E., Haimovitz, K., Ballweber, C., Dweck, C., & Popović, Z. (2014). Brain points: A growth mindset incentive structure boosts persistence in an educational game. *Proceedings of the SIGCHI Conference on Human Factors in Computing Systems*, 3339–3348. https://doi.org/10.1145/2556288.2557157

Pelánek, R. (2015). Metrics for evaluation of student models. *JEDM - Journal of Educational Data Mining*, 7(2), 1–19.

Pennycook, G., Ross, R. M., Koehler, D. J., & Fugelsang, J. A. (2017). Dunning-Kruger effects in reasoning: Theoretical implications of the failure to recognize incompetence. *Psychonomic Bulletin & Review*, 24(6), 1774–1784. https://doi.org/10.3758/s13423-017-1242-7

Picard, R. R., & Cook, R. D. (1984). Cross-validation of regression models. *Journal of the American Statistical Association*, 79(387), 575–583. https://doi.org/10.1080/01621459.1984.10478083

Powers, D. M. (2011). Evaluation: From precision, recall and F-measure to ROC, Informedness, Markedness and correlation. *Journal of Machine Learning Technologies*, 2(1), 37–63.

R Core Team. (2013). *R: A language and environment for statistical computing*.

Rico-Juan, J. R., Gallego, A.-J., & Calvo-Zaragoza, J. (2019). Automatic detection of inconsistencies between numerical scores and textual feedback in peer-assessment processes with machine learning. *Computers & Education*, 140, 103609. https://doi.org/10.1016/j.compedu.2019.103609

Robertson, J. (2011). The educational affordances of blogs for self-directed learning. *Computers & Education*, 57(2), 1628–1644. https://doi.org/10.1016/j.compedu.2011.03.003

Romero, C., Ventura, S., & García, E. (2008). Data mining in course management systems: Moodle case study and tutorial. *Computers & Education*, 51(1), 368–384. https://doi.org/10.1016/j.compedu.2007.05.016

Sanyal, D., Bosch, N., & Paquette, L. (2020). Feature selection metrics: Similarities, differences, and characteristics of the selected models. *Proceedings of the 13th International Conference on Educational Data Mining (EDM 2020)*, 212–223.

Schaffer, C. (1994). A conservation law for generalization performance. In W. W. Cohen & H. Hirsh (Eds.), *Proceedings of the Eleventh International Conference on Machine Learning* (pp. 259–265). Morgan Kaufmann. https://doi.org/10.1016/B978-1-55860-335-6.50039-8

Schmidt, J. A., Shumow, L., & Kackar-Cam, H. Z. (2017). Does mindset intervention predict students' daily experience in classrooms? A comparison of seventh and ninth graders' trajectories. *Journal of Youth and Adolescence*, 46(3), 582–602. https://doi.org/10.1007/s10964-016-0489-z

Selim, H. M. (2007). Critical success factors for e-learning acceptance: Confirmatory factor models. *Computers & Education*, 49(2), 396–413. https://doi.org/10.1016/j.compedu.2005.09.004

Sisk, V. F., Burgoyne, A. P., Sun, J., Butler, J. L., & Macnamara, B. N. (2018). To what extent and under which circumstances are growth mind-sets important to academic achievement? Two meta-analyses. *Psychological Science*, 29(4), 549–571. https://doi.org/10.1177/0956797617739704

Stephens, N. M., Fryberg, S. A., Markus, H. R., Johnson, C. S., & Covarrubias, R. (2012). Unseen disadvantage: How American universities' focus on independence undermines the academic performance of first-generation college students. *Journal of Personality and Social Psychology*, *102*(6), 1178–1197. https://doi.org/10.1037/a0027143

Warren, F., Mason-Apps, E., Hoskins, S., Devonshire, V., & Chanvin, M. (2019). The relationship between implicit theories of intelligence, attainment and socio-demographic factors in a UK sample of primary school children. *British Educational Research Journal*, *45*(4), 736–754. https://doi.org/10.1002/berj.3523

Wolpert, D. H., & Macready, W. G. (1997). No free lunch theorems for optimization. *IEEE Transactions on Evolutionary Computation*, *1*(1), 67–82. https://doi.org/10.1109/4235.585893

Wright, S. J. (2015). Coordinate descent algorithms. *Mathematical Programming*, *151*(1), 3–34. https://doi.org/10.1007/s10107-015-0892-3

Yeager, D. S., Hanselman, P., Walton, G. M., Murray, J. S., Crosnoe, R., Muller, C., Tipton, E., Schneider, B., Hulleman, C. S., Hinojosa, C. P., Paunesku, D., Romero, C., Flint, K., Roberts, A., Trott, J., Iachan, R., Buontempo, J., Yang, S. M., Carvalho, C. M., … Dweck, C. S. (2019). A national experiment reveals where a growth mindset improves achievement. *Nature*, 1–6. https://doi.org/10.1038/s41586-019-1466-y

Yeager, D. S., Henderson, M. D., D'Mello, S. K., Paunesku, D., Walton, G. M., Spitzer, B. J., & Duckworth, A. L. (2014). Boring but important: A self-transcendent purpose for learning fosters academic self-regulation. *Journal of Personality and Social Psychology*, *107*(4), 559–580. https://doi.org/10.1037/a0037637

Yeager, D. S., Romero, C., Paunesku, D., Hulleman, C. S., Schneider, B., Hinojosa, C., Lee, H. Y., O'Brien, J., Flint, K., Roberts, A., Trott, J., Greene, D., Walton, G. M., & Dweck, C. S. (2016). Using design thinking to improve psychological interventions: The case of the growth mindset during the transition to high school. *Journal of Educational Psychology*, *108*(3), 374–391. https://doi.org/10.1037/edu0000098

# Appendix A: Feature importance for all variables

In this appendix we provide a complete list of feature importance values. This also serves as a list of all variables included in the models.

**Table A1. Feature importance (mean absolute Shapley value) for all variables included in models 1 and 2. Variables are shown in decreasing order of model 2 feature importance. Most variables were obtained via student survey, and are thus self-reports.**

| Variable | Absolute effect on prediction | | | |
| --- | --- | --- | --- | --- |
| | **Model 1** | | **Model 2** | |
| | *M* | *SD* | *M* | *SD* |
| Pre-experiment GPA | 0.184 | 0.163 | 0.037 | 0.028 |
| Blocked navigation count | 0.016 | 0.010 | 0.019 | 0.036 |
| Race/ethnicity | 0.021 | 0.026 | 0.012 | 0.005 |
| Reason for learning | 0.005 | 0.005 | 0.012 | 0.007 |
| Expectations of success in high school math (time 2) | 0.041 | 0.025 | 0.008 | 0.006 |
| Distraction during intervention (time 1) | 0.002 | 0.004 | 0.008 | 0.003 |
| Self-reported typical grades | 0.106 | 0.076 | 0.007 | 0.011 |
| School culture regarding effort in classes | 0.010 | 0.014 | 0.007 | 0.008 |
| Other students worked quietly on the intervention (time 1) | 0.006 | 0.005 | 0.007 | 0.011 |
| Math teacher is respectful | 0.002 | 0.002 | 0.007 | 0.009 |
| Stress | 0.018 | 0.013 | 0.006 | 0.002 |
| Math interest | 0.004 | 0.004 | 0.006 | 0.002 |
| Highest level of parental education | 0.024 | 0.018 | 0.005 | 0.005 |
| Other students worked quietly on the intervention (time 2) | 0.005 | 0.004 | 0.005 | 0.009 |
| Teachers are respectful | 0.003 | 0.004 | 0.005 | 0.006 |
| Fixed mindset (time 2) | 0.021 | 0.014 | 0.004 | 0.002 |
| Math teacher explains importance of content | 0.014 | 0.015 | 0.004 | 0.003 |
| Challenge-seeking in math | 0.004 | 0.002 | 0.004 | 0.002 |
| Stress about high school classes | 0.003 | 0.003 | 0.004 | 0.002 |
| Math class is busy/efficient (time 2) | 0.003 | 0.003 | 0.004 | 0.002 |
| Math anxiety | 0.013 | 0.009 | 0.003 | 0.003 |
| Hard vs. easy exercises chosen on a worksheet | 0.006 | 0.009 | 0.003 | 0.002 |
| Technical difficulties during intervention (time 2) | 0.002 | 0.003 | 0.003 | 0.005 |
| Expectations of success in high school math (time 1) | 0.024 | 0.014 | 0.002 | 0.003 |
| International baccalaureate/gifted and talented | 0.017 | 0.013 | 0.002 | 0.001 |
| Fixed mindset (time 1) | 0.012 | 0.011 | 0.002 | 0.002 |
| Understanding what makes life meaningful | 0.006 | 0.007 | 0.002 | 0.002 |
| Math teacher explains things clearly | 0.005 | 0.007 | 0.002 | 0.001 |
| Sense of belonging in high school | 0.003 | 0.004 | 0.002 | 0.001 |
| Goal is to avoid looking dumb in classes | 0.003 | 0.003 | 0.002 | 0.002 |
| Free or reduced-price lunch eligibility | 0.072 | 0.041 | 0.001 | 0.000 |
| Gender | 0.037 | 0.013 | 0.001 | 0.001 |
| First year freshman | 0.022 | 0.020 | 0.001 | 0.000 |
| Special education status | 0.015 | 0.012 | 0.001 | 0.001 |
| Teachers are fair | 0.013 | 0.009 | 0.001 | 0.001 |
| Distraction during intervention (time 2) | 0.011 | 0.013 | 0.001 | 0.001 |
| Effort in school is seen as not cool | 0.009 | 0.007 | 0.001 | 0.001 |
| Student behavior is out of control in math class | 0.006 | 0.006 | 0.001 | 0.001 |
| Working hard in school feels meaningful | 0.006 | 0.004 | 0.001 | 0.001 |
| Technical difficulties during intervention (time 1) | 0.005 | 0.007 | 0.001 | 0.001 |
| Bad math grades attributed to low math ability | 0.004 | 0.007 | 0.001 | 0.001 |
| Trust in math teacher (time 1) | 0.003 | 0.004 | 0.001 | 0.001 |
| Trying really hard implies lack of skill | 0.002 | 0.003 | 0.001 | 0.001 |
| Trust in math teacher (time 2) | 0.002 | 0.002 | 0.001 | 0.001 |
| Math class is busy/efficient (time 1) | 0.001 | 0.002 | 0.001 | 0.001 |
| English language learner | 0.009 | 0.008 | 0.000 | 0.000 |
| Audio was on during intervention | 0.009 | 0.004 | 0.000 | 0.000 |
| School is safe | 0.009 | 0.009 | 0.000 | 0.000 |
| Math teacher explains things multiple ways | 0.002 | 0.003 | 0.000 | 0.001 |
| Math teacher believes everyone can be good at math | 0.001 | 0.002 | 0.000 | 0.000 |
| One racial group is not more important to my identity than another | 0.000 | 0.001 | 0.000 | 0.000 |

# Appendix B: Comparisons of models and cross-validation strategies

In this appendix we compare several of the key analysis choices made in this paper to alternatives.

First, we compared accuracy of model 1 and model 2 when using a linear regression model instead of an XGBoost model. For model 1, linear regression yielded $r = .295$ (mean across all cross-validation folds), versus $r = .327$ for XGBoost. For model 2, linear regression yielded $r = .065$, versus $r = .094$ for XGBoost. These results indicate that XGBoost captured between predictor and outcome variables slightly more effectively, but that the advantages gained by non-linearity and complex moderators was small. This aligns with the XGBoost feature important graphs, which show that there were some non-linear relationships and some interactions between variables, but for the majority of students a linear relationship was a close approximation.

Second, we compared the accuracy of model 1 without adjusting regularization hyperparameters. Model accuracy was lower ($r = .239$ versus $.327$) and accuracy on training data (i.e., without cross-validation) was $r = .999$. These results indicate that the model was severely over-fit, and highlighting the necessity of regularization.

Third, we computed the accuracy of model 1 and 2 (both XGBoost) while varying the number of schools in the training data using a leave-several-schools-out cross-validation approach. In the leave-several-schools-out approach we randomly selected from 1 to 50 schools to use as training data, and tested on the rest. We repeated the process 10 times for each number of randomly-selected schools and averaged the accuracies. This analysis is in contrast to the leave-one-school-out approach, where 61 of 62 schools were used for training. Results in Figure B1 below indicate that reducing the size of the training data reduced accuracy. Models have not reached a clear plateau, and it is possible that additional data collection might further improve accuracy. One possible explanation is that the search space of possible non-linear and complex relationships between variables is quite large, and a correspondingly large amount of data is required for machine learning approaches to successfully uncover these relationships.

Fourth, we repeated analysis of model 2 twice with a split sample (31 schools randomly selected for each half of the data). This analysis was intended to explore the stability of model 2 across data subsamples, given that effect sizes were small for model 2. We computed feature importance values for each split subsample and correlated them to determine how consisted the patterns learned by model 2 were. Feature importance values correlated $r = .575$, indicating substantial (though not total) replication across subsamples.

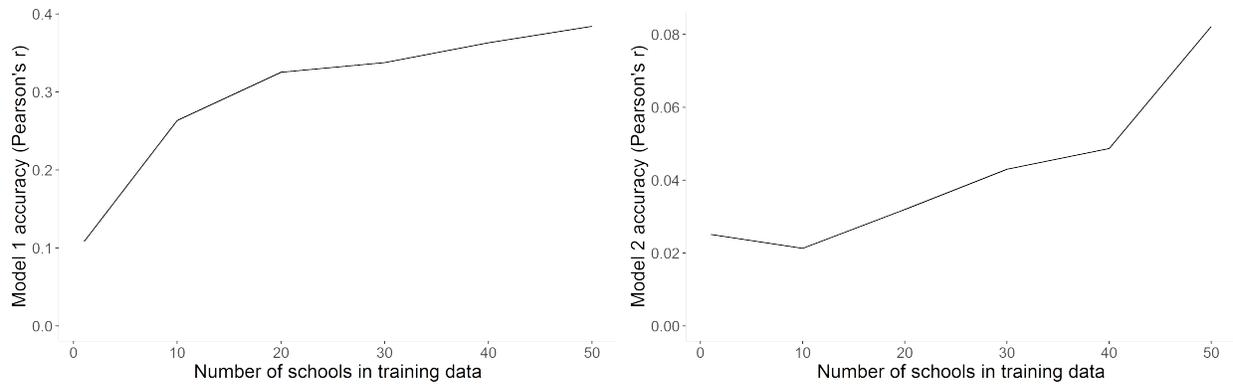

**Figure B1. Model accuracy as a function of training data size for model 1 (left) and model 2 (right).**